\documentclass
[aps,pre,showpacs,showkeys,superscriptaddress,twocolumn]{revtex4}%
\usepackage{amssymb}%
\usepackage{amsmath}%
\usepackage{amsfonts}%
\usepackage{graphicx}
\def\be{\begin{equation}}
\def\ee{\end{equation}}
\def\beq{\begin{eqnarray}}
\def\eeq{\end{eqnarray}}

\begin{document}
\title{ Exact entropy of dimer coverings for a class of lattices in 
three or more dimensions}
\author{Deepak Dhar}
\affiliation{Department of Theoretical Physics, Tata Institute of Fundamental Research,
Homi Bhabha Road, Mumbai 400005, India}
\author{Samarth Chandra}
\affiliation{Department of Theoretical Physics, Tata Institute of Fundamental Research,
Homi Bhabha Road, Mumbai 400005, India}
\date{\today}

\begin{abstract}

 We construct a class of lattices in three and higher 
dimensions for which the number of dimer coverings can be determined 
exactly using elementary arguments.  These lattices are a generalization 
of the two-dimensional kagome lattice, and the method also works for 
graphs without translational symmetry.  The partition function for dimer 
coverings on these lattices can be determined also for a class of 
assignments of different activities to different edges.


\end{abstract}

\maketitle

The dimer model is a well-known problem in classical lattice statistics. 
The Ising model can be reformulated as a dimer model on a modified graph 
\cite{fisher}. The dimer model is a simple, non-trivial but analytically 
tractable model which provides a starting point for the studies of 
systems of non-spherical molecules with hard-core interactions 
\cite{review}.  The latter can show a variety of geometrical phase 
transitions e.g. in liquid crystals \cite{liquidcrystals}. Recently 
there has been a lot of interest in quantum dimer models, to describe 
the ground states of frustrated quantum antiferromagnets, and some exotic 
phases expected in assemblies of hard core bosons on pyrochlore lattices 
\cite{theory}.

  Kasteleyn's  pioneering solution of the problem for 
general two-dimensional 
planar lattices \cite{kasteleyn}, led to a lot of follow-up work 
on the two-dimensional problem \cite{review}, but the problem in higher 
dimensions is much less studied. For non-zero fraction of sites not 
covered by dimers, it is known that there is no phase transition as a 
function of the density of dimers \cite{lieb}. Huse et al. have given 
non-rigorous arguments that the dimer-dimer orientational correlations 
have a power-law decay in all dimensions on bipartite lattices 
\cite{huse}. For graphs made of corner sharing triangles, where each 
vertex has a coordination number four, the entropy of dimer coverings 
per site can be determined exactly in any dimension \cite{misguich}.  
Huang et al studied the dimer problem in 3-dimensions, where the 
lattice is a stack of 2-d planes, but the problem is not fully 
3-dimensional as the dimers were confined to lie in 
planes \cite{huang}.  Priezzhev calculated exactly the number of a {\it 
subset} of all dimer configurations on the cubic lattice 
\cite{priezzhev}. For some anisotropic dimer models in three dimensions, 
the phase transition point can be determined exactly \cite{nagle}, but 
not the total entropy.

Elser \cite{elser} provided a simple argument to show that entropy per 
site of dimer covering of the kagome lattice (Fig. 1a) has a very 
simple 
form, and equals $ \frac{1}{3} \log 2$. This argument was extended 
to arbitrary graphs made of corner-sharing triangles by Misguich et al 
\cite{misguich}.  This problem was investigated again recently by Wang 
and Wu \cite{wangwu}.  The arguments of Misguich et al are valid only 
for graphs in which each vertex has coordination number $4$, and belongs 
to exactly two triangles. In this paper, we present an argument that is 
similar to the Elser's, but is valid for graphs in which the 
coordination number of vertices is not restricted to $4$.  Our arguments 
are applicable to a large class of lattices with arbitrary dimension, 
and also to graphs without translational invariance.  Some examples of 
lattices where our method can be used are shown in Fig. 1.

For simplicity of presentation, we first describe our arguments as 
applied to the kagome lattice (shown in Fig. 1a), and discuss the 
generalizations later. We select a subset of the sites of the lattice. 
We will call these the red sites, and denote the set by ${\mathcal R}$. 
For the kagome lattice, we choose the red sites to be the sites of one 
of the three sublattices original lattice (Fig. 1a).  For any dimer 
covering ${\cal C}$ of the lattice, to each red site $s$ of the lattice, 
we assign a discrete variable $\sigma_s$. This variable takes only two 
possible values: $ \sigma_s$ is $+1$ if the other vertex of the dimer 
covering $s$ is above $s$, and $-1$ if it is below.

 We now arbitrarily choose a value of $\sigma_s$ for each site $s$ in 
${\mathcal R}$. Let us denote this set by $\{\sigma_s\}$, and ask how 
many dimer configurations  are consistent with 
$\{\sigma_s\}$. We note that if $\sigma_s = +1$, then the two bonds 
connecting the site $s$ to sites below it would not be used in dimer 
coverings. Then we can safely delete these bonds, and consider dimer 
covering of the remaining graph. Similarly, if $\sigma_s = -1$, we 
remove two bonds connecting $s$ to sites above. If we do this for all 
sites of the red sublattice, the full lattice breaks up into a set of 
mutually disconnected chains of the type shown in Fig. 2.

 It is easy to see that each chain can be covered by dimers in at most 
one way.  For example, in Fig. 2, for the top chain, starting from left, 
we see that site $1$ has to be matched with site $3$, and then site $2$ 
has to matched with $4$. And so on.

 It is convenient to adopt the boundary conditions that a site on the 
{\it right boundary} can be either covered by a dimer, or left uncovered 
\cite{boundary}.   Then there is exactly one dimer covering consistent 
with any given set $\{\sigma_s\}$.  The number of red sites is $N/3$, 
and so the number of different  possible choices of 
$\{\sigma_s\}$ is $2^{N/3}$.   
This is also the total number of dimer coverings.

We define entropy per site $C$ by the relation that number of dimer 
coverings increases as $\exp( CN )$, for a lattice
\begin{widetext}

\begin{figure}
\begin{center}
\begin{tabular}{cc}
   \resizebox{60mm}{!}{\includegraphics[height=50mm, 
width=60mm]{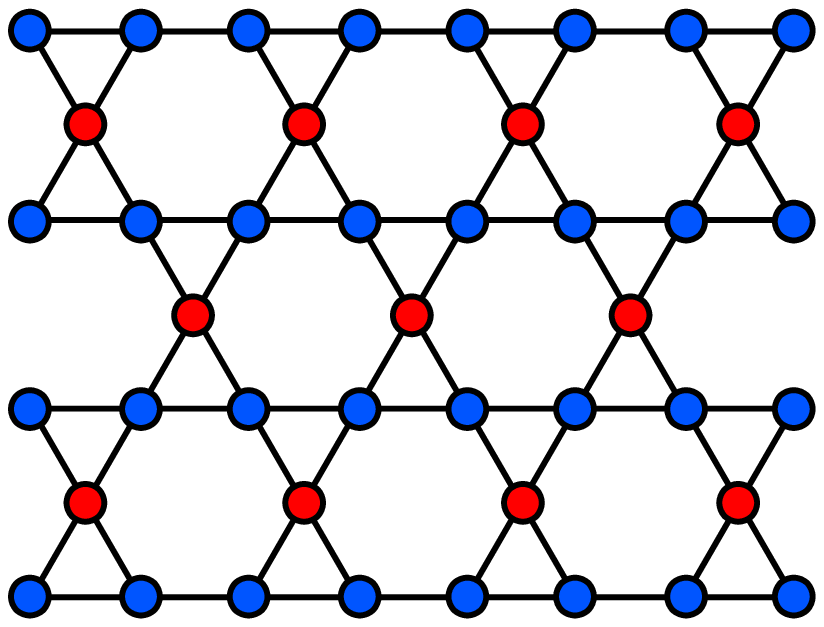}} &    
\resizebox{60mm}{!}{\includegraphics[height=50mm, 
width=50mm]{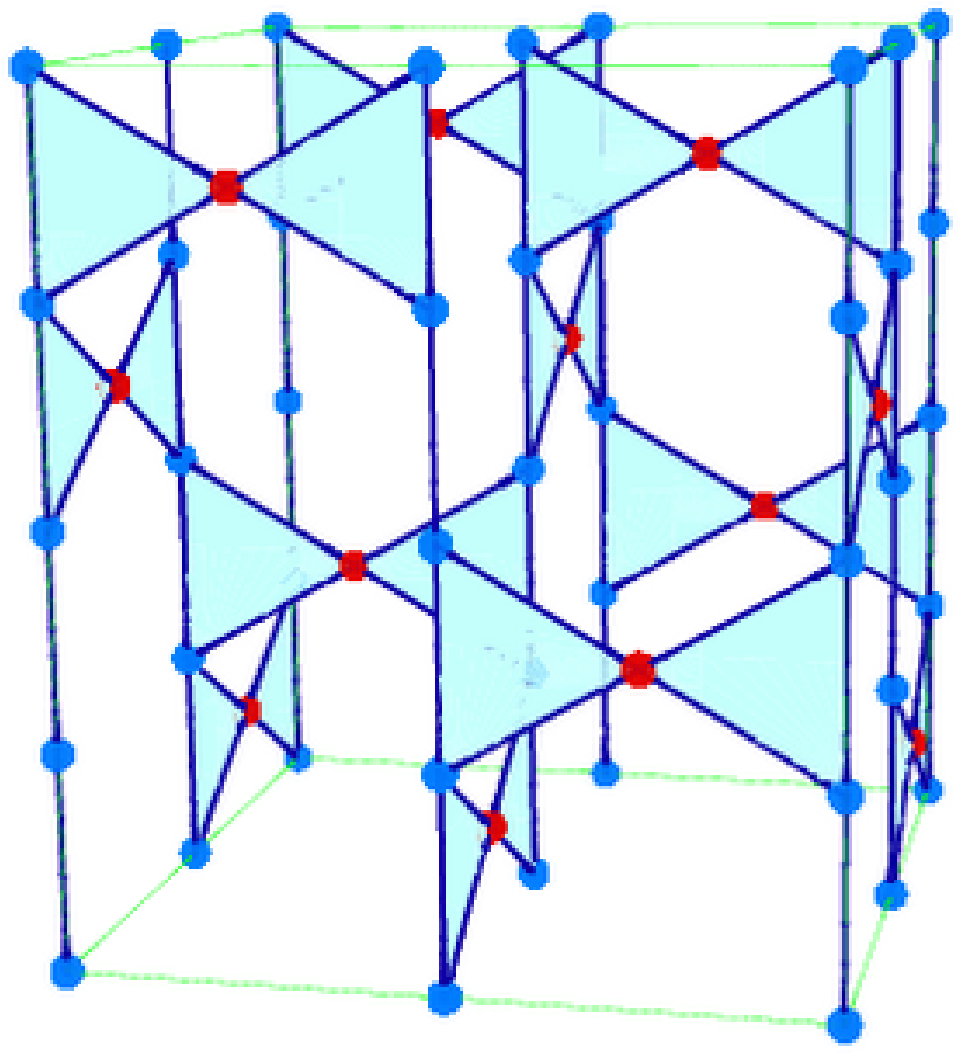}} \\
(a) & (b) \\
\resizebox{60mm}{!}{\includegraphics[height=50mm, 
width=50mm]{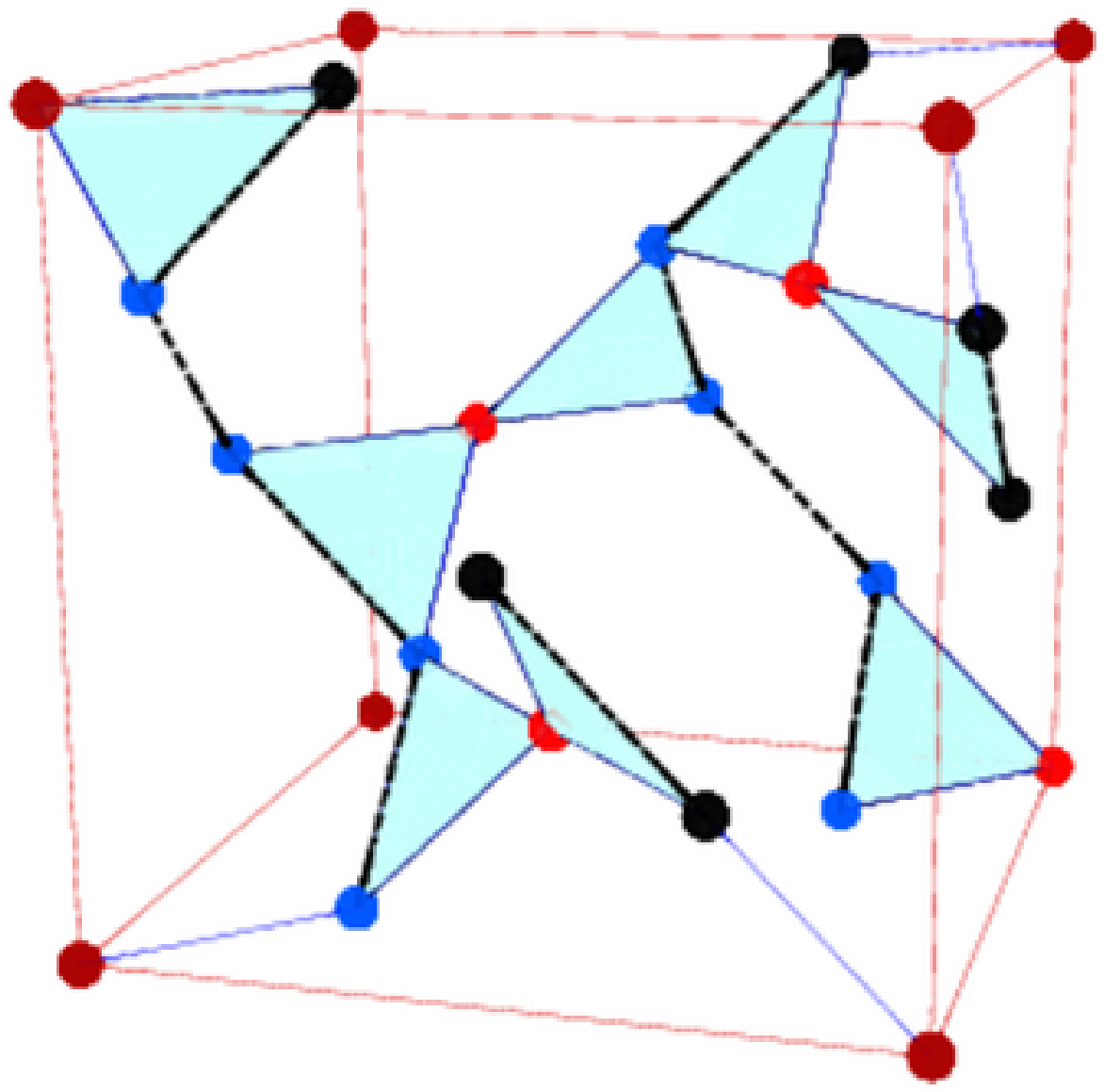}} &    
\resizebox{60mm}{!}{\includegraphics[height=50mm, 
width=50mm]{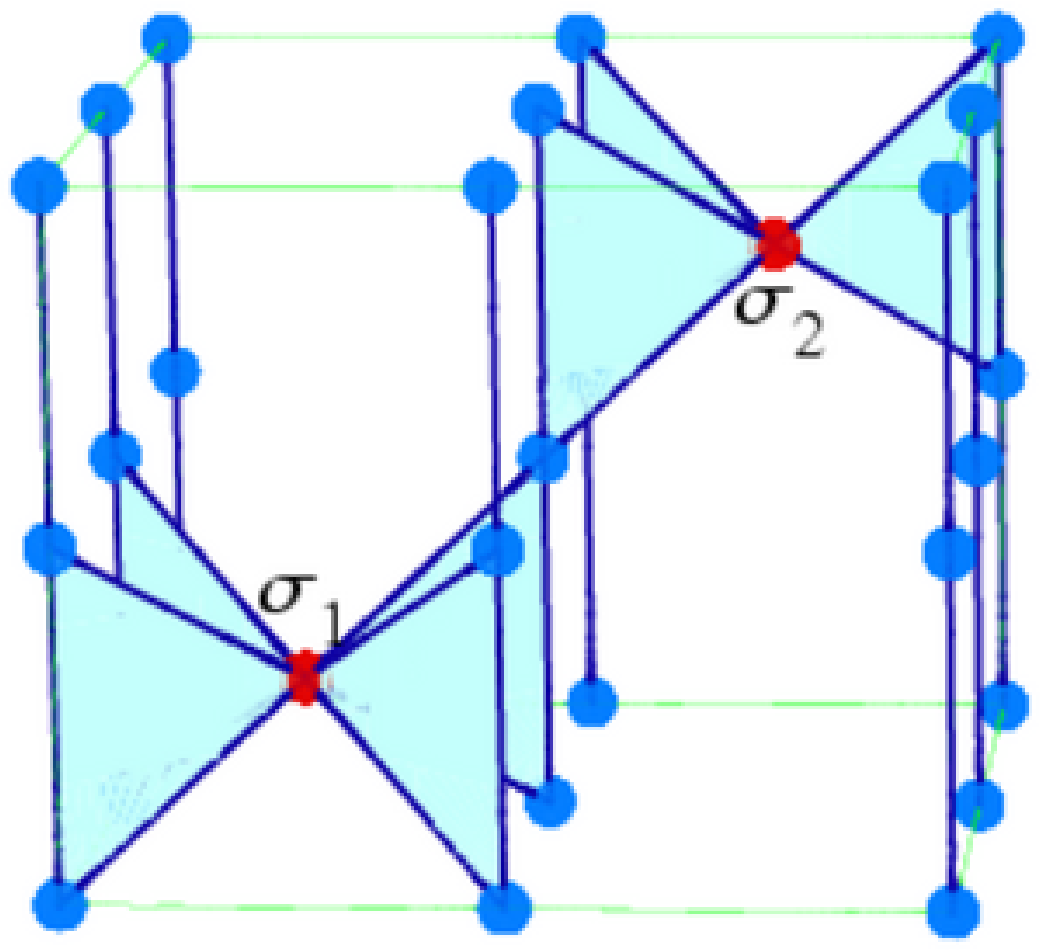}}\\
(c) & (d)  
\end{tabular}

\caption{Some examples of lattices for which the exact entropy is 
calculated in this paper. (a) The kagome lattice. (b) A 
three-dimensional lattice with corner-sharing triangles (c) The unit 
cell of the $Na_4 Ir_3 O_8$ lattice. There are 12 Iridium atoms per unit 
cell (shown in red or blue). Sodium and oxygen atoms are not  shown. 
Atoms belonging to other unit cells are shown in dark red or 
black. (d) Unit cell of a lattice obtained by decoration 
procedure discussed in text.  } 
\end{center} 
\label{fig1} 
\end{figure}

\end{widetext}
 with $N$ sites, for 
large $N$.  Then  for the kagome lattice, 
\begin{equation}
C = \frac{1}{3} \log 2.
\end{equation}

The above argument can be generalized to more general connected graphs 
made of corner-sharing triangles.  Consider, for example, the lattices 
shown in Fig. 1b and 1c. To construct the lattice of Fig. 1b we start 
with a simple cubic lattice with only the vertical edges present. For 
each pair of neighbouring vertical chains we add a red site in 
between every fourth pair of opposing edges and join it to the ends of 
the edges. The opposing edges in between whom we add these additional 
sites are displaced with respect to each other so that no two triangles 
share an edge, as shown in Fig. 1b. 

The unit cell of a
hyperkagome lattice is shown in Fig. 1c. There are experimental materials like the 
sodium iridate ($Na_4 
Ir_3O_8$) where this structure, known as the GGG structure, is realized 
in nature \cite{footnote}. To obtain this 
lattice we start from the pyrochlore 
lattice which consists of corner sharing tetragons, their centers 
forming a diamond lattice. The hyperkagome lattice in Fig. 1c is obtained 
by removing one site from each tetragon to leave a lattice of corner 
sharing triangles.

 For these cases, we choose the red sites as shown in Figs. 1b and 1c, 
and 
as in the kagome case, to each red site assign a binary variable 
$\sigma$, specifying whether the dimer covering that site belongs to one 
or the other of the two triangles meeting at that point.  Then, as 
before, we ask how many different dimer coverings are consistent with an 
arbitrary choice of $\{\sigma_s\}$.

\begin{figure} 
\begin{center} 
\includegraphics[scale=.7]{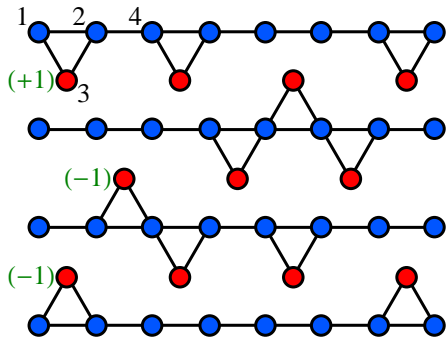} 
\caption{ On  deleting two edges from each red site, the kagome lattice 
breaks into  disconnected  chains.  The values of the $\sigma$ variables 
at some of the red sites 
are indicated by numbers in parantheses.  } 
\label{fig2} 
\end{center} 
\end{figure}

If now, from each red site, we delete two edges that are not consistent 
with the given choice, the lattice breaks up into mutually disconnected 
chains of single edges and triangles similar to Fig. 2. These chains 
have a "backbone" of bonds
in the $z$-direction in Fig. 1b, and a zig-zag array of roughly 
vertical lines in Fig. 1c (shown by black dashed lines). If we assume 
free boundary conditions on one (say the top) face, there is a unique 
dimer covering of the lattice consistent with any aribrary choice of 
$\{\sigma_s\}$.  For large $N$, the number of red sites is $N/3$ (up to 
surface correction terms), and the number of dimer coverings is 
$2^{N/3}$. Hence, for both the $3$-dimensional lattices shown in Fig. 1b 
and 1c, we get

\begin{equation}
C = \frac{1}{3} \log 2
\end{equation}

For graphs in which each vertex is shared by two triangles, this result 
was  obtained earlier by Misguich et al 
\cite{misguich}.
But our  argument is extended easily to graphs where more than 
two 
triangles meet at a single vertex.  Consider a graph formed by the 
vertices and edges of a collection of non-intersecting (possibly 
non-straight) lines, which may be imagined as embedded in a 
$d$-dimensional lattice $(d \geq 2)$.  We add a number of additional 
vertices not lying on any line, and call these additional vertices red 
vertices.  The red vertices can be labelled by integers $1, 2, \ldots 
N_r$. For the $j$th red vertex, we choose a positive integer $r_j \geq 
1$, choose $r_j$ different edges from the original lines, and join both 
ends of each edge to the red vertex $j$ by additional edges. Then the 
$j$th red vertex is the shared corner vetex of $r_j$ triangles, and its 
coordination number is $2 r_j$. We ensure that the edges selected do not 
already have some vertex in common, and that no edge of the original 
lines is selected more than once, so that no edge belongs to more than 
one triangle.  This defines the graph ${\mathcal L}$ for which we want 
to calculate the number of possible dimer covers.

An example of such a graph is shown in Fig. 1d. The lattice consists of 
vertices of a simple cubic lattice, labelled by coordinates 
$(m_1,m_2,m_3)$, where $m_1, m_2, m_3 $ are integers. We start with 
edges only along the $z$-direction. The lattice graph consists of 
disconnected vertical $1$-dimensional chains. We add extra connections 
to this graph to make it $3$-dimensional as follows: we pick an 
elementary cube of the lattice, add an extra vertex at the center of the 
cube, and connect it to two or more of the vertical edges of the cube by 
triangles (an example is shown in Fig. 1d). We choose a unit cell of 
period $p_1, p_2, p_3$ where $p_1, p_2, p_3$ are integers $> 1$.  (In 
the case shown in Fig. 1d, $p_1 = p_2 = p_3 = 2$.) We choose some extra 
cubes within the unit cell to connect this way, and then repeat this 
pattern to get a translationally invariant lattice. We ensure that these 
`decorations' are selected such that the resulting graph is connected, 
and that no edge belongs to more than one triangle.

    In the example shown in Fig. 1d, we chose two diagonally opposite 
cubes within the unit cell, and add a site each at the center of each 
cube. The site with coordinates $(1/2, 1/2, 1/2)$ is connected to all 
its four neighboring vertical edges, but we connect the site at $(3/2, 
3/2, 3/2)$ to only three of its neighboring vertical edges.

 Then, as before, at each red vertex $j$, we define a variable 
$\sigma_j$ which takes $r_j$ possible values. There is a unique dimer 
cover of ${\mathcal L}$ consistent with a 
particular arbitrarily chosen set of values $\{\sigma_s\}$.    It 
follows that the total number of dimer covers of this graph $ \Omega $ 
is given by

\begin{equation}
\Omega =    \prod_{j = 1}^{N_r}  r_j
\end{equation}

In the Fig. 1d, we have two red vertices per unit cell, and $r_1 =4$, 
and $ r_2 =3$.  There are $10$ sites per cell. Hence the entropy per 
site for the lattice shown in Fig. 1d is $\frac{1}{10} \log 12$.

It is easy to see that the arguments can be trivially extended to other 
choices of decorations, and also to higher dimensional lattices.

Our treatment can also be extended to the case where different edges are 
associated with different activities.  This is illustrated most easily 
for the kagome lattice.  Let us associate activity $z_1$ with a 
horizontal edge of the lattice (Fig. 1a). Let us assume that the 
activity for a non-horizontal edge is $z_2$ if it belongs to an 
up-pointing triangle, and $z_3$ if it belongs to a down pointing 
triangle. Then, for a particular $\{\sigma_s\}$ having $n$ values $+1$, 
the weight of the unique dimer cover is easily seen to $z_2^{n} z_3^{N/3 
-n } z_1^{N/6}$. Summing over all choices of $\{\sigma\}$, we get total 
partition function 
$\Omega(z_1, z_2, z_3) = ( z_2 + z_3)^{N/3} z_1^{N/6}$.  This answer 
differs from that obtained by Wang and Wu \cite{wangwu}, as our choice 
of edge weights differs from theirs.

It is easy to see that this same method works for the higher dimensional 
graphs (e.g. for the lattice shown in Fig. 1d), so long as  (a) 
the two edges at a red vertex belonging to the same triangle always have 
the same weight, and (b) all edges not having a red end vertex have the 
same weight.

For example, consider the lattice shown in Fig. 1d. In the unit cell 
shown in Fig. 1d there are seven triangles. Let the non-vertical edges 
of the respective triangles have the activities $z_1, z_2, \ldots z_4$ 
for the bonds in the $4$ triangles meeting at $(1/2,1/2,1/2)$, $z_5, 
z_6, z_7$ for the bonds meeting at the site $(3/2, 3/2, 3/2)$ and $z_0$ 
be the activity for the vertical edges. Then the free energy per site is 
$(1/10) \log[( z_1 + z_2 + z_3 + z_4)(z_5 + z_6 + z_7) z_0^3]$.

Note that the fact that $\sigma$ variables are completely uncorrelated, 
suggests that orientational correlations between dimers on these 
lattices may be short-ranged. In fact, for the kagome lattice, the 
dimer-dimer orientational correlation function has been shown to be zero 
for all separations $>2$ lattice spacings \cite{wangwu}.

We should emphasize that though the $\{\sigma_s\}$ variables are 
local, independent variables, that uniquely specify the allowed dimer 
configurations, the 
procedure of finding the dimer configuration corresponding to a given 
$\{\sigma_s\}$ is nontrivial, and non-local. If we change $\sigma_s$ at 
only one red site, it changes the orientations of dimers very far along 
the two affected vertical chains.

 We thank Kedar Damle for very insighful discussions, and pointing us to 
literature on experimental realizations of hyper-kagome lattices, and 
Drs. M. Barma, K. S. Krishnan, S. Maiti, J. Radhakrishnan, A. K. Raina 
and V. Tripathi for their critical reading of the manuscript. We 
thank Dr. G. Misguich for bringing refs. \cite{elser,misguich} to our 
notice.  This 
research has been supported in part by the Indo-French Center for 
Advanced Research under the project number 3404-2. SC thanks C.S.I.R., 
India for financial support through the Shyama Prasad Mukherjee 
fellowship.


\begin{thebibliography}{9}
\bibitem{boundary} Other choices of boundary conditions are possible. 
The bulk entropy does not depend on these details.
\end{thebibliography}
\end{document}